\documentclass[preprint,12pt,authoryear]{elsarticle}

\usepackage{amssymb}

\journal{Planetary and Space Science}

\begin{document}

\begin{frontmatter}



\title{Physical properties of Taurid meteoroids of various sizes}


\author{Ji\v{r}\'\i\ Borovi\v{c}ka and Pavel Spurn\'y}

\address{Astronomical Institute of the Czech Academy of Sciences, Fri\v{c}ova 298, CZ-25165 Ond\v{r}ejov,
Czech Republic}

\begin{abstract}
The origin of the Taurid complex is still debated. In addition to comet 2P/Encke, various asteroids
were proposed to be members of the complex and thus possible parent bodies of Taurid meteoroids. 
Studies of physical properties of Taurid meteoroids can provide constrains on their source. We used a well defined
orbital sample of 16 Taurid fireballs with detailed radiometric light
curves. The sample represented meteoroids of initial masses from 8
grams to 650 kg (diameters 1 -- 70 cm). The semi-empirical fragmentation
model was used to study their atmospheric fragmentation and derive
strength distribution within the meteoroids. It was found that small meteoroids are stronger than large ones. 
When considering Taurid material as a whole, the majority has
a very low strength of less than 0.01 MPa and a density less than 1000 kg m$^{-3}$. 
The low strength material exists 
mostly as large bodies ($>10$ cm). 
If encountered in smaller bodies, it forms only a minor part.
Stronger materials up to 0.3 MPa exist in Taurids as well, but the stronger material the rarer it is.
Strong material forms small inclusions in large bodies 
or exists as small (cm-sized) separate bodies. These properties strongly suggest cometary origin of Taurids.  
\end{abstract}




\end{frontmatter}


\section{Introduction}
\label{intro}

Taurids is an annual meteor shower characterized by a low but long activity, lasting for more
than two months \citep{Jennbook}. Taurid meteoroids have short orbital periods of about 3.4 years and low
inclinations. The comet with the shortest period, 2P/Encke, is a plausible parent body but it was proposed
that 2P/Encke is just one fragment of a much larger comet, which disrupted tens of thousands years ago \citep{Clube}
and formed a whole complex.
A number of asteroids was proposed to be members of the Taurid complex and thus parent bodies of Taurid
meteoroids \citep[see][and references therein]{chapter8}.

In some years, Taurid activity is enhanced and the shower is rich in fireballs. It was proposed that the enhanced
activity is caused by a swarm of meteoroids trapped in the 7:2 resonance with Jupiter \citep{QJRAS}. 
\citet{Spurny} proved that the enhanced activity in 2015 was caused by a well-defined orbital structure,
with meteoroids indeed being in the 7:2 resonance. This structure, called the new branch, was unexpectedly 
active also in 2018 \citep{poster8}, meaning that the resonant swarm is more extended than supposed by \citet{QJRAS}.

This paper is devoted to physical properties of Taurid meteoroids. The knowledge of physical properties
is important for understanding the origin of the Taurid complex. There is also non-negligible impact risk
with members of the Taurid complex. The new branch, which intersects the Earth's orbit, contains also
asteroids of sizes of the order of 100 m \citep{Spurny}. Understanding the structure of Taurid material 
can help to evaluate potential consequences of collision with such a large body.

Physical studies of Taurids performed so far gave mixed results. The estimates of bulk densities are dependent
on the used model and vary from 400 kg m$^{-3}$ \citep{BellotRubio} or 1600 kg m$^{-3}$ \citep{Babadzhanov}
to 2300 -- 2800 kg m$^{-3}$ \citep{Konovalova}. \citet{Konovalova} studied Taurid atmospheric fragmentation 
events
in detail and found them to occur under dynamic pressures of 0.05 -- 0.18 MPa. She also found significant lateral
velocities of fragments and concluded that fragmentations are explosive events. \citet{Matlovic} found
Taurids to be a heterogeneous population of meteoroids, which are cometary in nature but contain solid, possibly
carbonaceous inclusions. The derived {\em mineralogical} densities varied from 1300 to 2500 kg m$^{-3}$
(bulk densities are expected to be lower due to a porosity). Fragmentation strengths were in the range
0.02 -- 0.10 MPa.

Since the orbits of Taurids fall in the region occupied by many Near Earth Asteroids, studies of Taurids can be
affected by contamination of meteoroids with different origin. Both \citet{Brown} and \citet{Madiedo} reported
Taurid fireballs which possibly dropped meteorites and were therefore quite strong. However, the orbits of those fireballs
suggest that they may not belong to Taurids at all. Here we study 16 Taurids for which there is no doubt
that they really belong to the shower: 15 of them were members of the well-defined new branch (14 were
observed in 2015 and one in 2018), one was a regular Southern Taurid. Our sample contains meteoroids
of masses covering five orders of magnitude: from $10^{-2}$ kg to almost $10^3$ kg. \citet{Spurny} already noted 
that the largest meteoroids belong to the most fragile type IIIB when classified according to the PE criterion
(comparing end height with initial mass and speed and trajectory slope)
of \citet{PE}, while small meteoroids belong to type II and some of them even to the strongest type I.
In this paper we want to shed light on this interesting pattern by detailed fragmentation modeling.

\begin{figure}
\includegraphics[width=9cm]{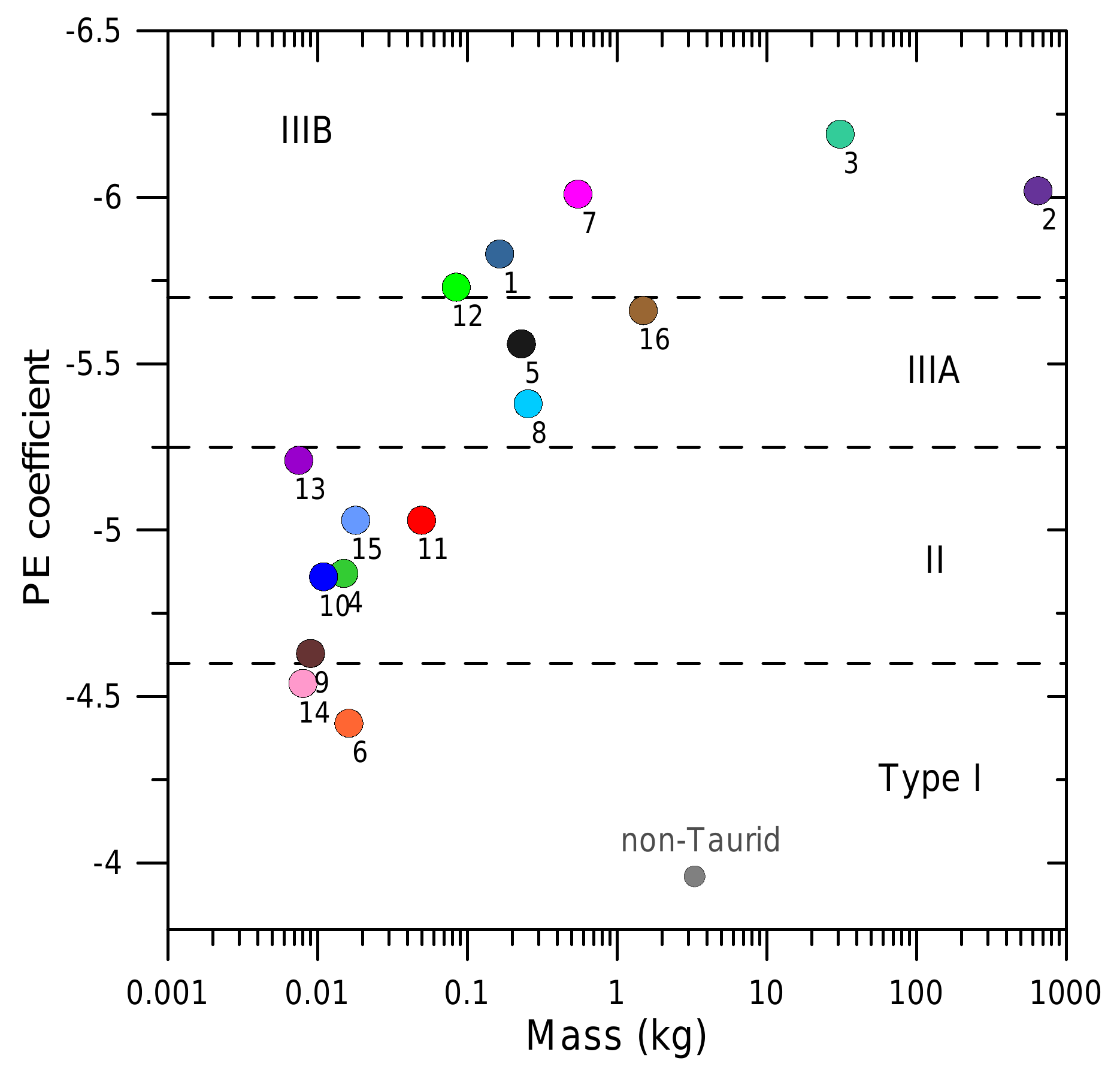}
\caption{PE coefficient as a function of mass for Taurid fireballs studied here. For fireball numbers see Table~\ref{table}.
One non-Taurid was added for comparison. 
Fireball type according to the PE coefficient is indicated.}
\label{PE}
\end{figure}

\section{Data and methods}

\subsection{Instrumentation}

The data analyzed here were obtained by the Digital Autonomous Fireball Observatories (DAFO)
in the scope of the European Fireball Network. The observatories have been introduced
in \citet{Spurny}. They provide digital all-sky images and radiometric light curves of fireballs.
Fireball trajectories, velocities, decelerations, and orbits are computed from multi-station photographs.
Velocity measurements are enabled by a LCD shutter interrupting the 35 s long exposures 16 times
per second. At the beginning of each second, one interruption is skipped, enabling the absolute time
of each interruption to be determined with the help of the radiometer. The radiometer measures the total
brightness of the sky with the frequency of 5000 Hz. The data are absolutely timed. Radiometers therefore
provide both absolute time and detailed fireball light curves. Radiometric light curves are calibrated using
photographic light curves, where fireball magnitudes can be determined by comparison with stars. 

\begin{figure}
\centering
\includegraphics[width=9cm]{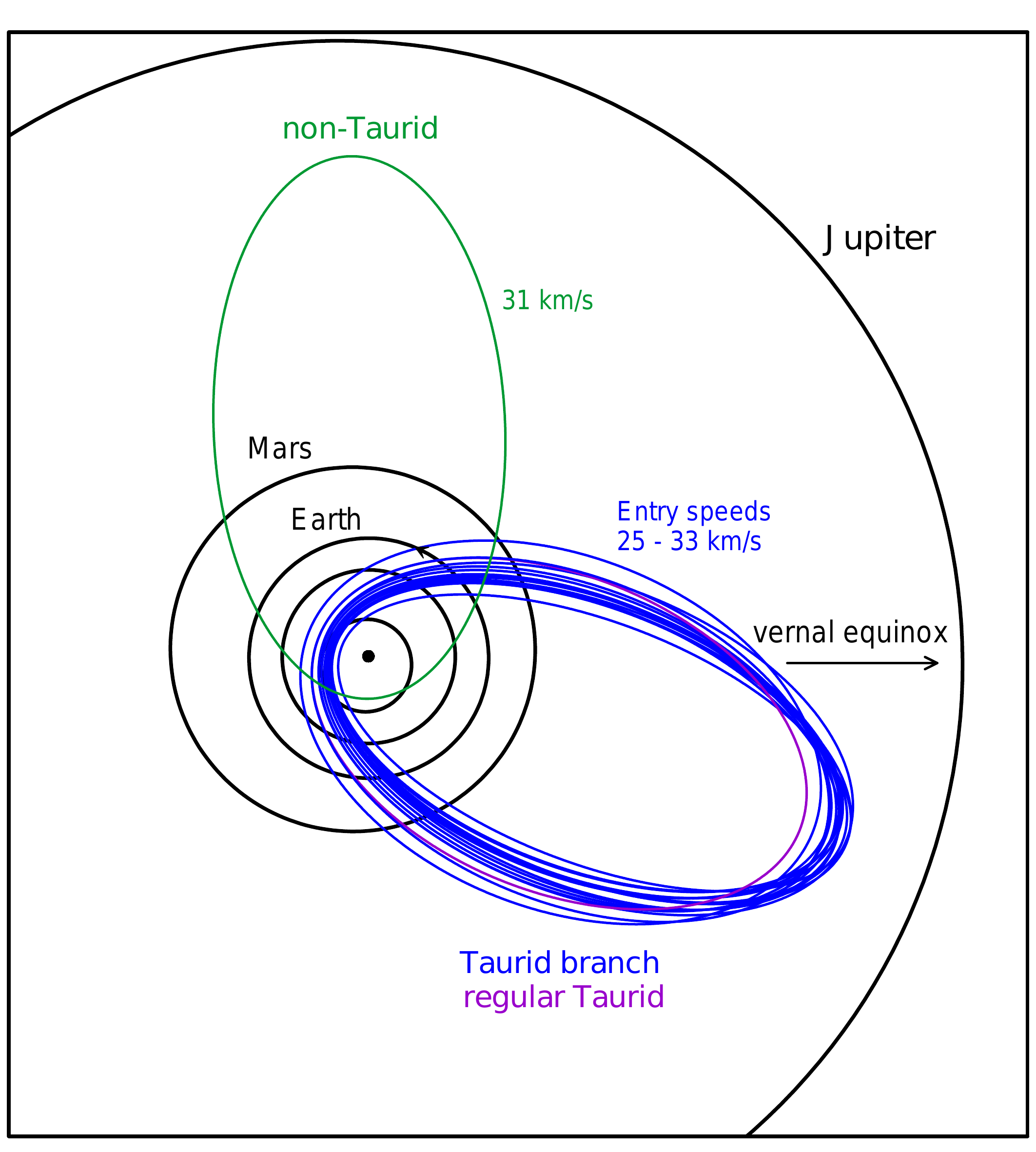}
\caption{Orbits of studied Taurids and one non-Taurid used for comparison.}
\label{orbits}
\end{figure}

\subsection{Data}

For the fragmentation analysis, 15 fireballs from the sample studied by \citet{Spurny} were selected.
The sample was supplemented by the brightest Taurid fireball observed in 2018, EN\,28118\_034715. The sample covers
all fireball types from I to IIIB. Figure~\ref{PE}, modified from \citet{Spurny}, shows the PE
coefficient as a function of meteoroid mass. 
The masses of the analyzed meteoroids ranged from 8 grams to 650 kg. 
The fact that larger meteoroids are more fragile is clearly visible in Fig.~\ref{PE}.

\begin{table*}
\caption{Selected meteoroid parameters derived from the fragmentation model}
\label{table}
\small
\begin{tabular}{lllrrllllll}
\hline
No. & Fireball code & Mass & Density & Size & $p_1$ & $p_{90\%}$ & $p_{\rm max}$ & \multicolumn{3}{c}{Fraction} \\
 & & kg & kg m$^{-3}$ & cm & MPa & MPa & MPa & Abla. & Ero. & Dust \\
\hline   
01& EN251015\_022301  &    0.165 & 1000. &   3.4& 0.008 & 0.025 &0.063 & 0.016 &  0.51 &  0.47  \\
02& EN311015\_180520  &    650.  &  500. &  67.7& 0.002 & 0.012 &0.19  & 0.001 &  0.98 &  0.02 \\ 
03& EN311015\_231302  &    31.   & 1400. &  17.4& 0.009 & 0.051 &0.19  & 0.003 &  0.95 &  0.05 \\ 
04& EN011115\_013625  &    0.015 & 1000. &   1.5& 0.070 & 0.085 &0.17  & 0.17 &  0.00 &  0.83 \\ 
05& EN021115\_022525  &    0.23  &  500. &   4.8& 0.020 & 0.044 &0.079 & 0.023 &  0.80 &  0.17 \\ 
06& EN021115\_232112  &    0.016 & 2000. &   1.3& 0.16  & 0.40  &0.40  & 0.60 &  0.00 &  0.40  \\
07& EN041115\_021111  &    0.55  & 1500. &   4.4& 0.015 & 0.051 &0.20  & 0.013 &  0.91 &  0.08 \\
08& EN041115\_203853  &    0.255 & 1000. &   3.9& 0.004 & 0.044 &0.13  & 0.09 &  0.66 &  0.25  \\
09& EN051115\_023102  &    0.009 & 1000. &   1.3& 0.046 & 0.100 &0.16  & 0.31 &  0.40 &  0.29  \\
10& EN061115\_040629  &    0.011 &  500. &   1.7& 0.001 & 0.045 &0.075 & 0.26 &  0.30 &  0.44  \\
11& EN081115\_181258  &    0.050 &  200. &   3.9& 0.005 & 0.031 &0.073 & 0.18 &  0.41 &  0.41  \\
12& EN101115\_235401  &    0.085 &  600. &   3.2& 0.016 & 0.047 &0.083 & 0.04 &  0.48 &  0.48  \\
13& EN131115\_002058  &    0.008 &  250. &   1.9& 0.008 & 0.057 &0.057 & 0.14 &  0.24 &  0.62  \\
14& EN161115\_193458  &    0.008 &  600. &   1.5& 0.011 & 0.132 &0.15  & 0.57 &  0.13 &  0.30  \\
15& EN161115\_213048  &    0.018 & 1000. &   1.6& 0.031 & 0.059 &0.080 & 0.08 &  0.23 &  0.69 \\
16& EN281118\_034715  &    1.5   &  250. &  11.3& 0.005 & 0.035 &0.097 & 0.06 &  0.65 &  0.29  \\  
\it nT&\it EN270217\_023122  &\it    3.3  &\it 3400. &\it  6.1 &\it  0.11 &\it 2.8 &\it 3.5 &\it 0.57 &\it  0.26 &\it  0.17 \\
\hline
\end{tabular}
\end{table*}

Figure~\ref{orbits} shows the orbits of the modeled fireballs. There is no doubt that all 16 fireballs were
members of the Taurid shower. The fireball list is provided in Table~\ref{table}. Fireball no.\ 14 was a regular Southern
Taurid; other Taurids belonged to the new branch.
For the comparison of physical properties, we also included one non-Taurid,
EN\,270217\_023122, observed in February 2017, which had similar semimajor axis and eccentricity
as Taurids and its entry speed (31 km s$^{-1}$) was within the Taurid range (25 -- 33 km s$^{-1}$).
The meteoroid physical properties were, however, very different, as it is shown in Table~\ref{table}
and will be discussed later.

\begin{figure}
\includegraphics[width=6cm]{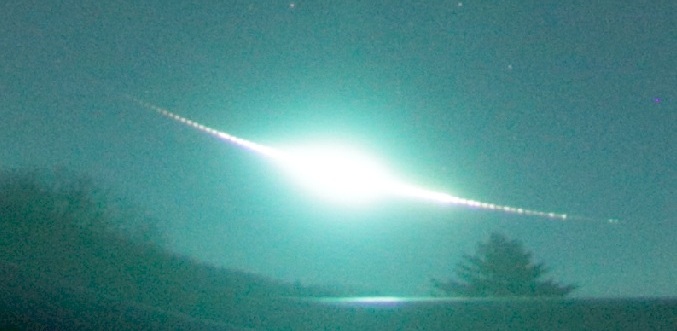} \hspace{0.5cm}
\includegraphics[width=6cm]{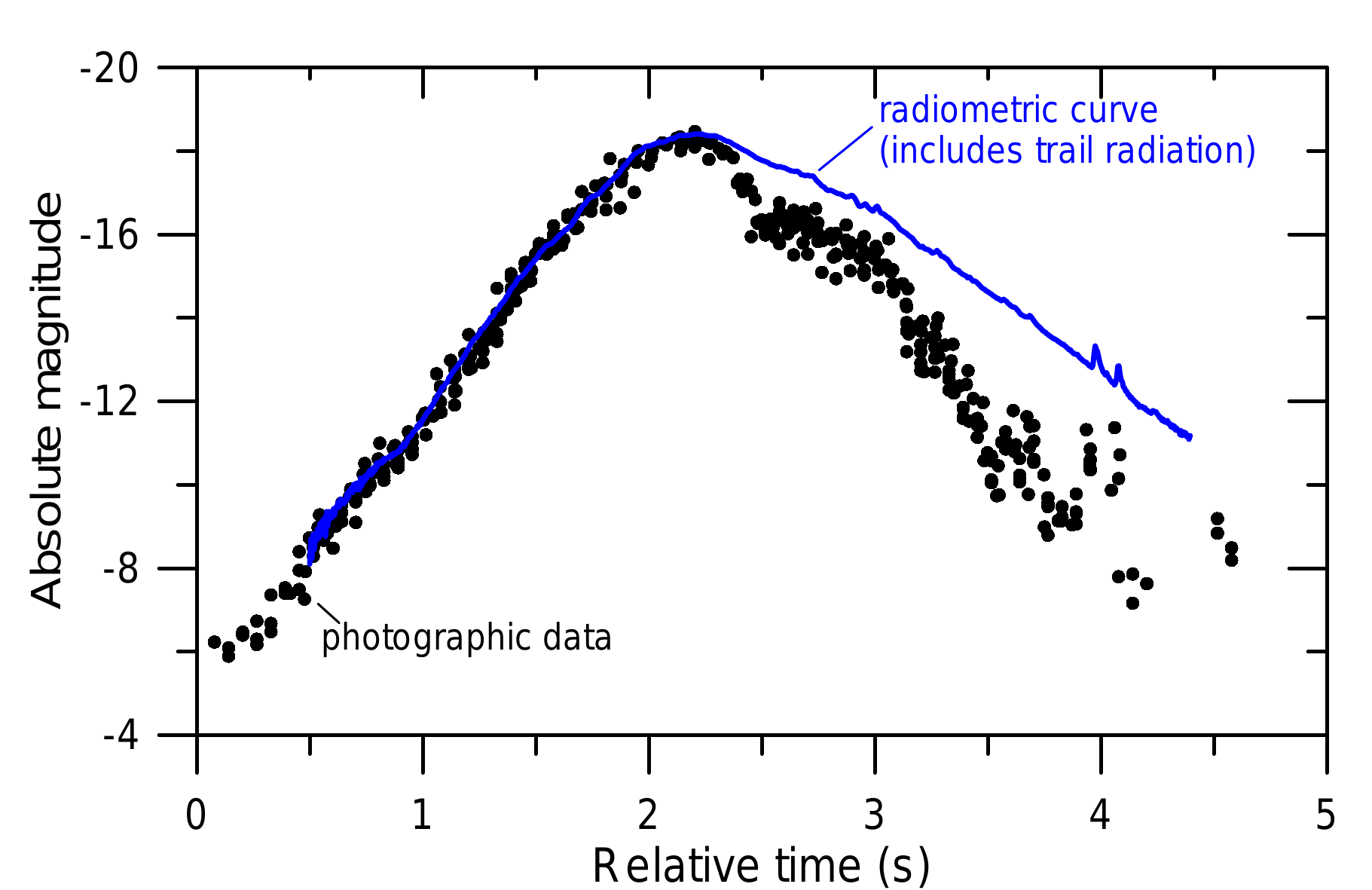}
\caption{Image and light curve of Taurid fireball no.\ 2 (EN\,311015\_180520).
Dots are photographic data 
(from eight different cameras), 
the curve is from a radiometer.}
\label{foto2}
\end{figure}

\begin{figure}
\includegraphics[width=4cm]{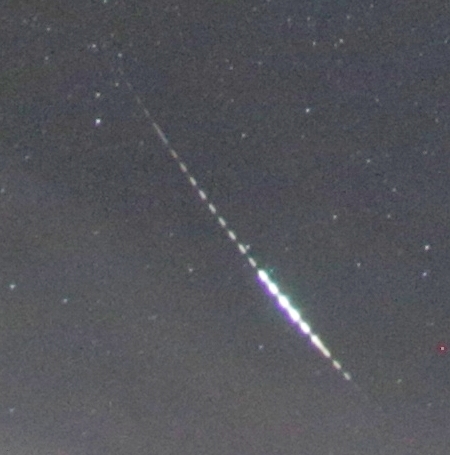} \hspace{2.5cm}
\includegraphics[width=6cm]{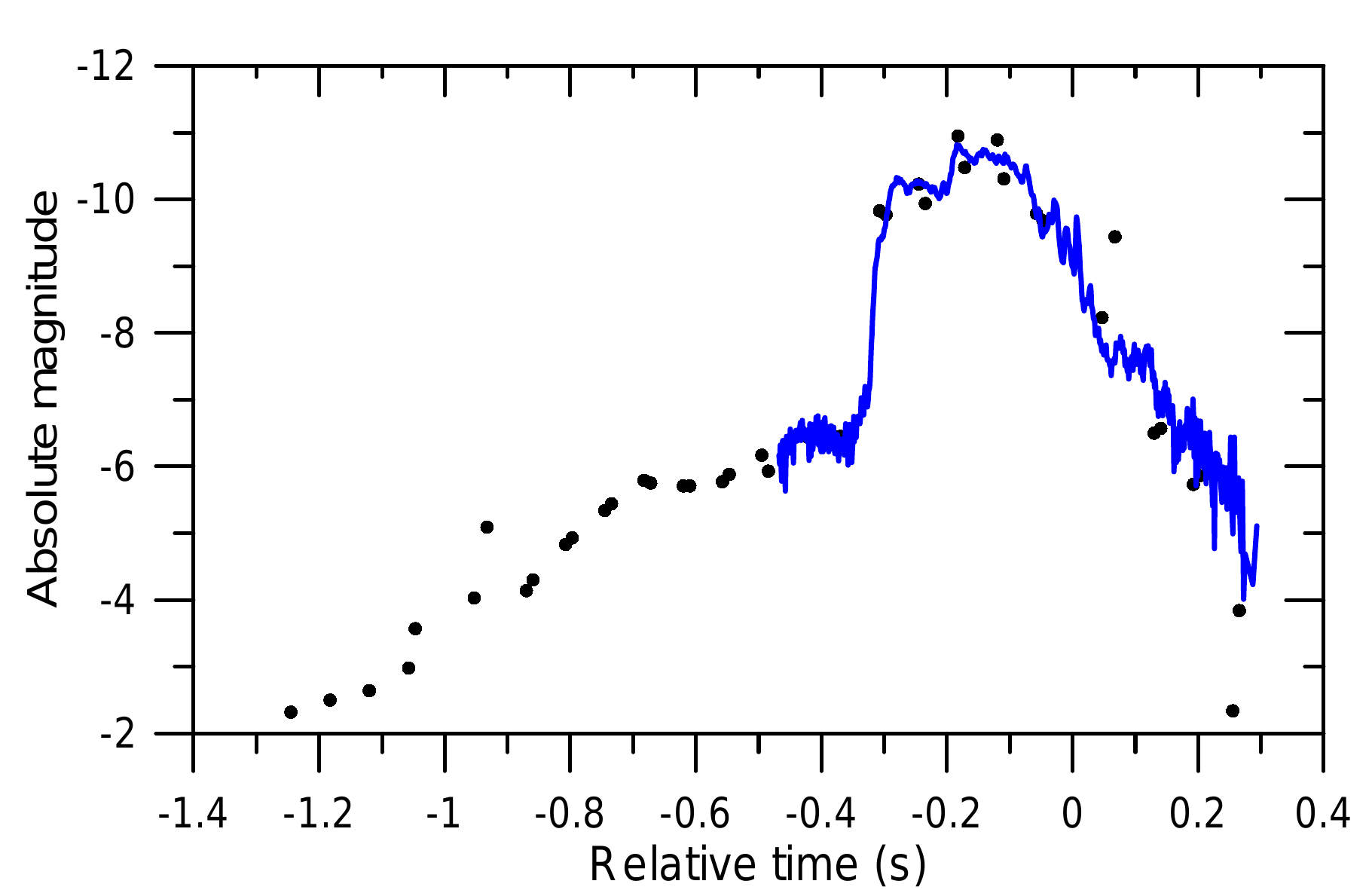}
\caption{Image and light curve of Taurid fireball no.\ 5 (EN\,021115\_022525)
Dots are photographic data
(from two different cameras), 
the curve is from a radiometer.}
\label{foto5}
\end{figure}

\begin{figure}
\includegraphics[width=6cm]{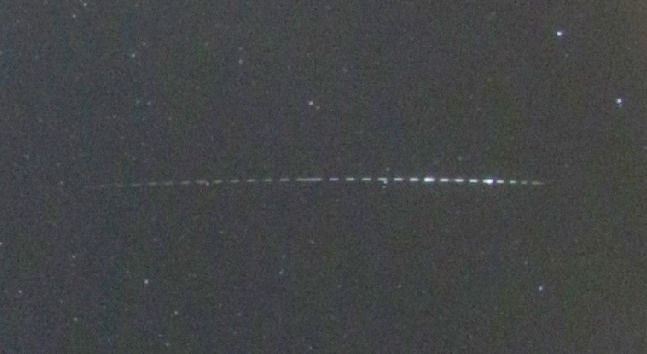} \hspace{0.5cm}
\includegraphics[width=6cm]{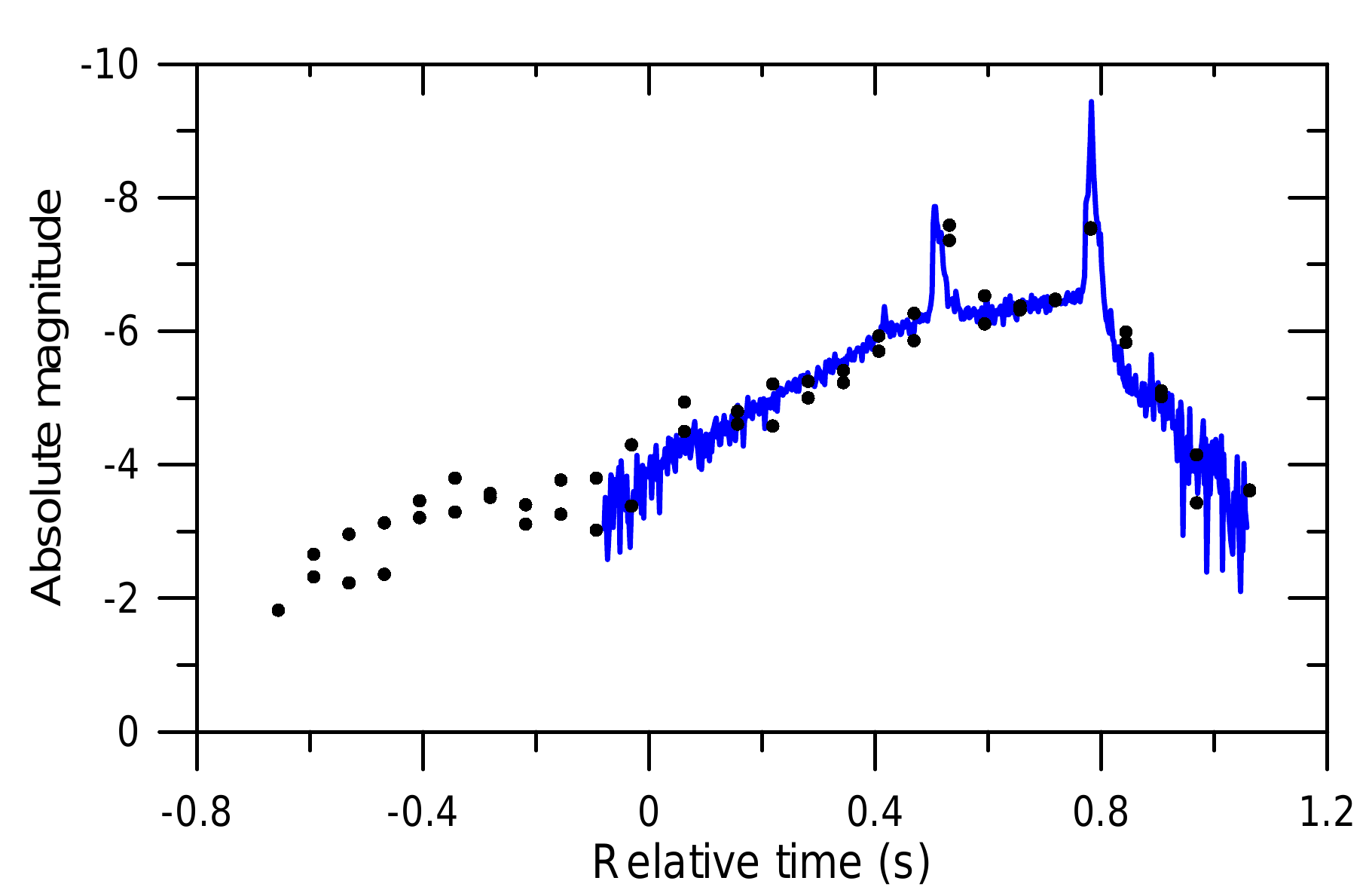}
\caption{Image and light curve of Taurid fireball no.\ 6 (EN\,021115\_232112).
Dots are photographic data
(from two different cameras), 
the curve is from a radiometer.}
\label{foto6}
\end{figure}

Figures~\ref{foto2}--\ref{foto6} shows images and light curves of three Taurids of various brightness.
The brightest observed fireball no.\ 2 had nearly symmetrical light curve and the maximal
brightness of $-18$ mag was reached already at a height of 81 km. There were also several short
flares toward the end of the fireball. The radiometric light curve contains also the radiation
of the stationary trail, which was very bright and formed around the position of the bolide maximum. The photographic
light curve describes only the moving meteor.
The medium brightness fireball no.\ 5 was characterized by a sudden increase of brightness by 4 magnitudes
(i.e.\ nearly 40$\times$) in the second half of the trajectory (at a height 78 km), which formed the beginning of a broad flare. 
Fireball no.\ 6 was relatively faint, except two short flares. This fireball penetrated deeper than the previous 
two, down the height of 48 km (the end height was 54 km for fireball no.\ 2 and 65 km for fireball no.\ 5). 

\subsection{Modeling}
\label{model}

\begin{figure}
\includegraphics[width=9cm]{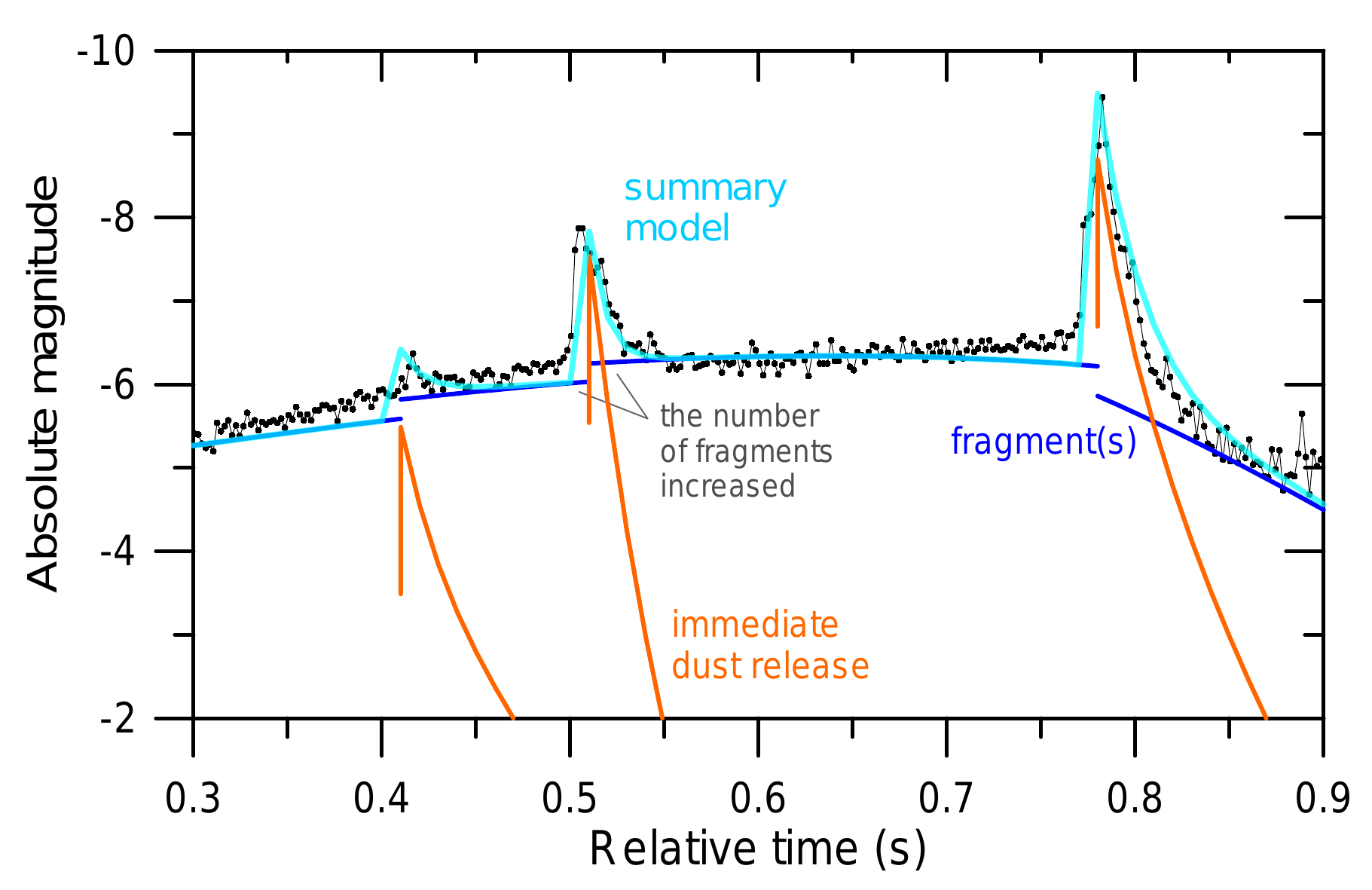} 
\caption{Fit of the light curve of Taurid fireball no.\ 6 demonstrating the explanation of short flares
by immediate dust releases.}
\label{model-dust}
\end{figure}

\begin{figure}
\includegraphics[width=9cm]{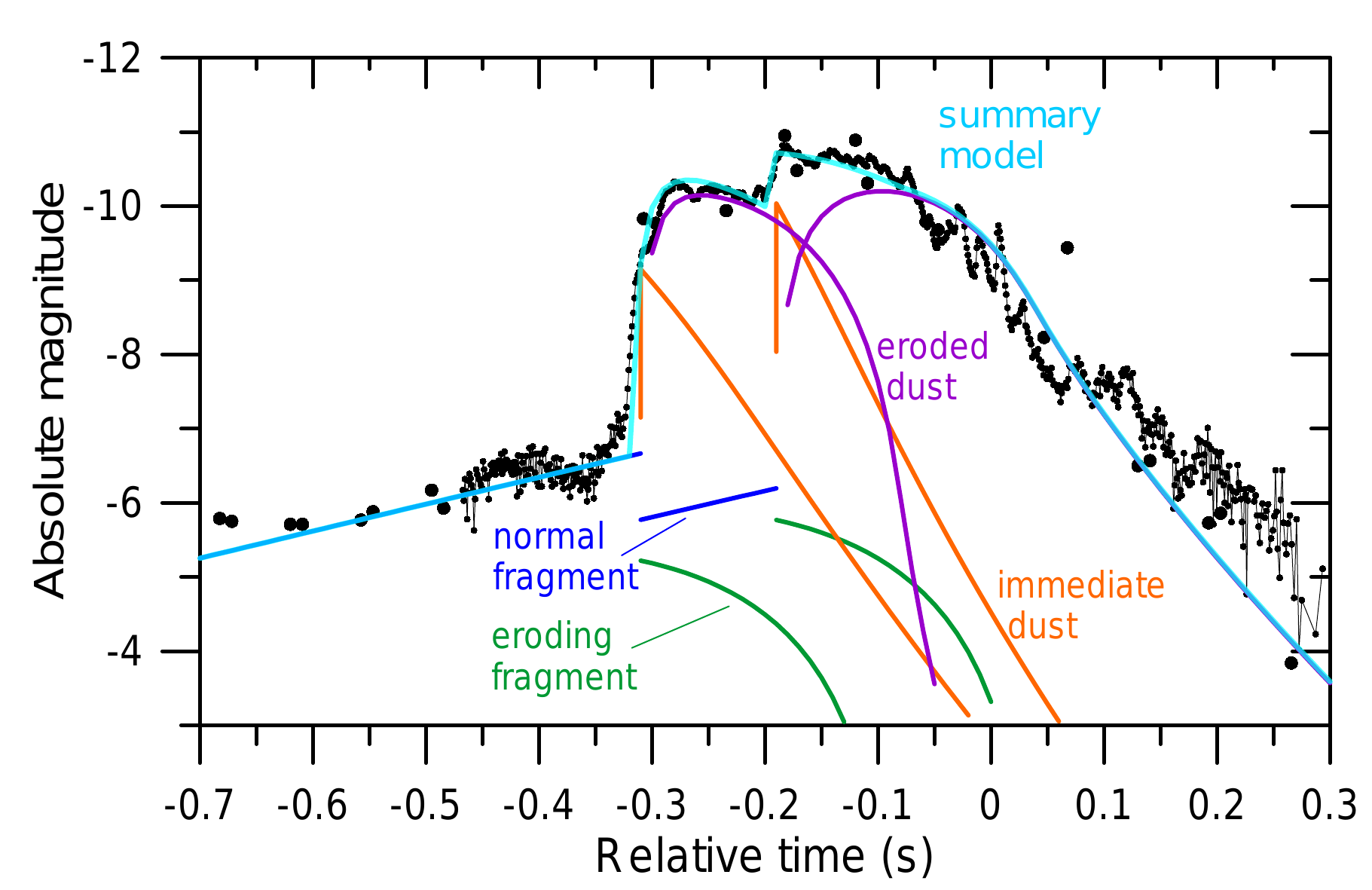} 
\caption{Fit of the light curve of Taurid fireball no.\ 5 demonstrating the dominating role of eroded dust in explanation 
of long flares.}
\label{model-erosion}
\end{figure}

The observed light curves and decelerations of fireballs were fitted by the semi-empirical fragmentation
model. The model was described in detail in \citet{Kosice} and \citet{chapter1}. Short flares were modeled by
an immediate release of dust, i.e.\ large number of small fragments, as it is illustrated in Fig.~\ref{model-dust}.
Longer flares were modeled with the help of the formalism of eroding fragments (Fig.~\ref{model-erosion}). Eroding
fragments are releasing dust from their surfaces over prolonged period of time. The rate of mass loss in form
of solid fragments is described by the erosion coefficient, analogous to the ablation coefficient 
(which describes the mass lost due to evaporation). A step-wise increase of brightness was modeled by
fragmentation into a number of macroscopic regular (non-eroding) fragments 
(Fig.~\ref{model-dust}). 
Usually, all fragments were assumed to have the same mass for simplicity.

The main parameters of the model were the initial mass and density of the meteoroid, the heights of fragmentations,
the mass of released dust, erosion fragments, and regular fragments at each fragmentation, the masses 
of dust grains, the densities of fragments, and the erosion coefficients.  The fixed parameters were the
intrinsic ablation coefficient 0.005 s$^2$km$^{-2}$ \citep[see][]{Ceplecha05}, 
the product of drag and shape coefficients, $\Gamma\!A=0.8$, dust grain
density 2000 kg m$^{-3}$, and the luminous efficiency dependent on velocity and mass in the same way as for the
Ko\v{s}ice meteoroid \citep{Kosice}.

We are mostly interested in the heights of fragmentations, which together with the known velocity at that point, $v$,
define the dynamic pressure $p=\rho v^2$. The atmospheric density $\rho$ was taken from the NRLMSISE-00 model
\citep{NRLMSISE}. We define that the dynamic pressure is equal to the strength of the meteoroid, namely the
strength of the part which was lost.  The lost part is defined as the mass before fragmentation minus the mass
of the largest regular fragment after fragmentation. Since multiple fragmentations were observed in all fireballs,
strength distribution within the meteoroids could be studied. The heights of the fragmentations were quite
obvious from the positions of flares on the light curve. The mass analysis is more model-dependent but the amplitudes 
and durations of flares are good indicators of the amount of lost mass.

Another parameter of interest is the meteoroid bulk density. It was primarily determined from the fireball brightness
before the first fragmentation. If the meteoroid mass is known (basically from the total radiated energy) and the ablation
coefficient and $\Gamma\!A$ are fixed, the brightness of the fireball, when it is still single body and the ablation
reached its steady-state, depends on
meteoroid density. Lower density means larger cross section and larger brightness. In some cases, low density
derived this way was confirmed by high deceleration. However, the used $\Gamma\!A=0.8$ assumes nearly spherical meteoroids.
If a meteoroid was highly non-spherical, the derived density will be not valid. 
We must therefore consider the individually derived
densities as rather uncertain. Note also that since the grain density was assumed to be 2000 kg m$^{-3}$ (for
presumably carbonaceous material), the highest possible meteoroid density was also 2000 kg m$^{-3}$.               

\section{Results}

The list of modeled fireballs and some  results of the model are given in Table~\ref{table}. Note that the
mass of the largest meteoroid (650 kg) is lower than 1300 kg given in \citet{Spurny} since 
only the radiation of the moving fragments was modeled. 
The radiation of the stationary trail, which was very intense in this case, cannot by explained 
by the fragmentation model and does not therefore enter to the energy budget.  

Apart from the trail, the fragmentation model was able to explain the light curves and decelerations of Taurid fireballs almost
perfectly. The only general problem seems to be terminal flares. 
Three fireballs (nos.\ 11, 12, and 15)
exhibited bright terminal flares while
the brightness before the flare was low and the deceleration also suggested meteoroid mass too low to produce the flare.
The discrepancy could be solved by adjusting some parameters 
(ablation coefficient, $\Gamma\!A$, or luminous efficiency) differently for the end of the 
fireball. The derived dynamic pressures at fragmentations were 
not affected by this problem.

For fireball no.\ 9, there was a problem to explain the increased deceleration after the first fragmentation
at the height of 72 km. Since the brightness after the flare did not increase in comparison with the situation 
before the flare, the deceleration could not be explained by a disruption into small fragments. We had to assume
that the velocity suddenly decreased by 0.4 km s$^{-1}$ at the fragmentation. This was the only fragmentation event in
our sample where a velocity change was introduced.

Deceleration was observed for all fireballs except no.\ 3, which was big and disintegrated quickly without leaving any observable 
fragments, which could be decelerated at lower heights. For most fireballs, the velocity at the end was 5 -- 10 km s$^{-1}$ lower
than at the beginning. In some cases, the velocity decreased significantly only after the last fragmentation,
in some cases it was decreasing along almost the whole trajectory. Fireballs nos.\ 1,  7, and 12 showed only minor decelerations
with velocity difference less than 2 km s$^{-1}$. On the other hand, fireball no.\ 6 decelerated from 32 to 12 km s$^{-1}$.
All these data were taken into account during the modeling.

\begin{figure}
\centering
\includegraphics[width=9.5cm]{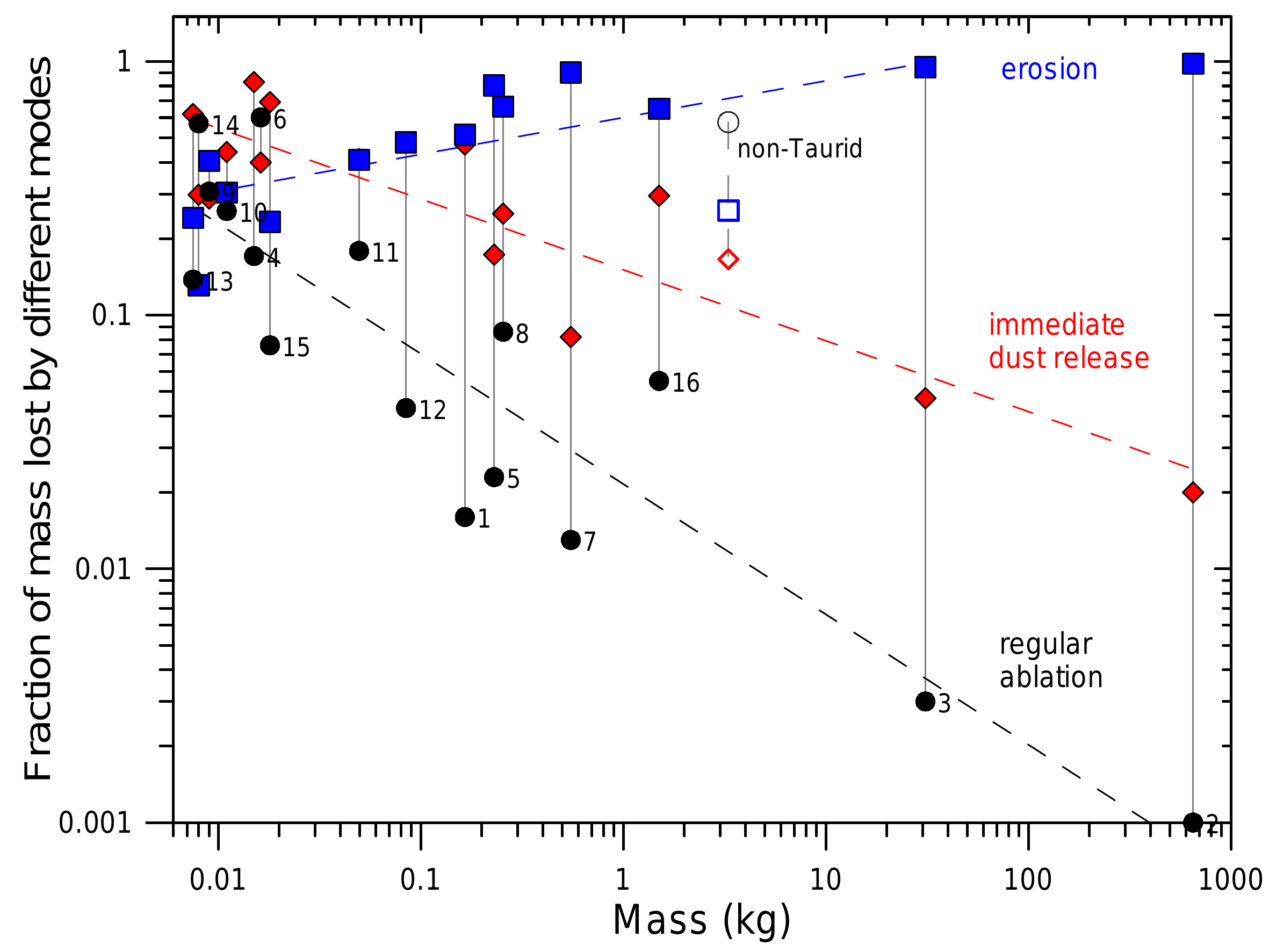}
\caption{Part of the meteoroid mass lost by regular ablation of macroscopic fragments (black circles), immediate release
of dust (red diamonds), and dust released gradually from eroding fragments (blue squares) as a function of initial mass.
For meteoroid numbers see Table~\ref{table}.
The same data for the non-Taurid meteoroid are also shown (empty symbols).}
\label{lossmode}
\end{figure}

\subsection{Modes of mass loss}

Figure~\ref{lossmode} shows the fraction of mass lost by regular ablation (i.e.\ evaporation from macroscopic fragments, not dust),
by erosion (i.e.\ in the form of dust from eroding fragments), and by suddenly released dust. There is a clear trend with mass.
Erosion was the dominant process for all Taurid meteoroids larger than 0.1 kg. For smaller meteoroids, immediate
dust release was increasingly important. Moreover, for two small meteoroids, ablation was the dominant 
process amounting for $\sim 60$\%
of lost mass (see also Table~\ref{table}). Any kind of fragmentation  was therefore not so important for them.
The non-Taurid did not follow the trend. Although having more than 3 kg, most mass was lost by regular ablation.

The masses of dust particles, both immediately released and eroded, ranged from $10^{-12}$ kg to $5\times10^{-6}$ kg.
It corresponds,  for the assumed density of 2000 kg m$^{-3}$, to sizes from about 10 $\mu$m to 1.5 mm. 
Meteoroids no.\ 3 and 6 contained only grains $\geq 100$ $\mu$m. The opposite case was meteoroid no.\ 12, which
contained only grains $\leq  100$ $\mu$m. The typical size range for other meteoroids was 20 -- 500 $\mu$m.

\subsection{Dynamic pressures}

\begin{figure}
\centering
\includegraphics[width=9.5cm]{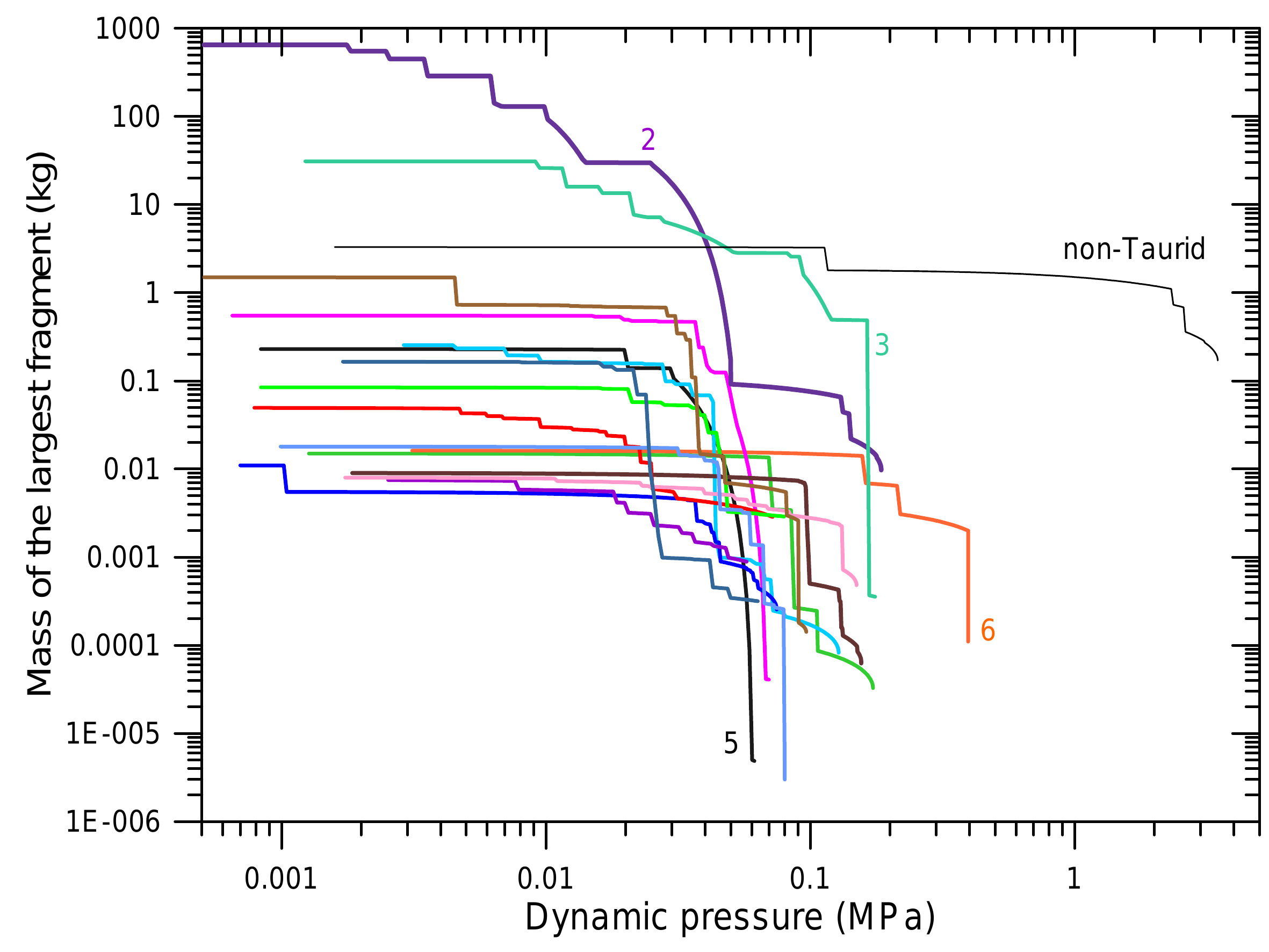}
\caption{Mass of the largest surviving fragment as a function of increased dynamic pressure for 16 Taurids and one non-Taurid.
Selected Taurids are identified by their numbers. For color legend see Fig.~\ref{press-mass-pct}.}
\label{press-mass}
\end{figure}

Figure~\ref{press-mass}  shows the mass loss of individual meteoroids as a function of increasing dynamic pressure
during the atmospheric entry.  Many meteoroids were subject to numerous fragmentation events. They
demonstrated themselves by multiple flares on the light curve. Although the largest meteoroid no.\ 2 
exhibited one very wide and bright flare, it was to be
modeled by subsequent release of five eroding fragments. On the other hand, in some
cases, e.g.\ meteoroids no.\ 5 and 6, there were only few (2--3) fragmentations. 

All Taurids were much weaker than the non-Taurid meteoroid. 
The non-Taurid reached dynamic pressures up to 3 MPa. 
All Taurids except one were destroyed before they 
reached 0.2 MPa. In fact, parts of only three Taurids approached pressures of about 0.15 MPa.
The largest of them was a 0.5 kg fragment of the $\sim 30$ kg meteoroid no.\ 3. About 20 g (only 0.003\% of the initial mass)
remained
from the huge meteoroid no.\ 2. But the small meteoroid no.\ 6 (14~g at that time) only started to fragment
at this pressure. Two parts of it, both of about 2 grams, survived almost to 0.4 MPa.

\begin{figure}
\centering
\includegraphics[width=7.5cm]{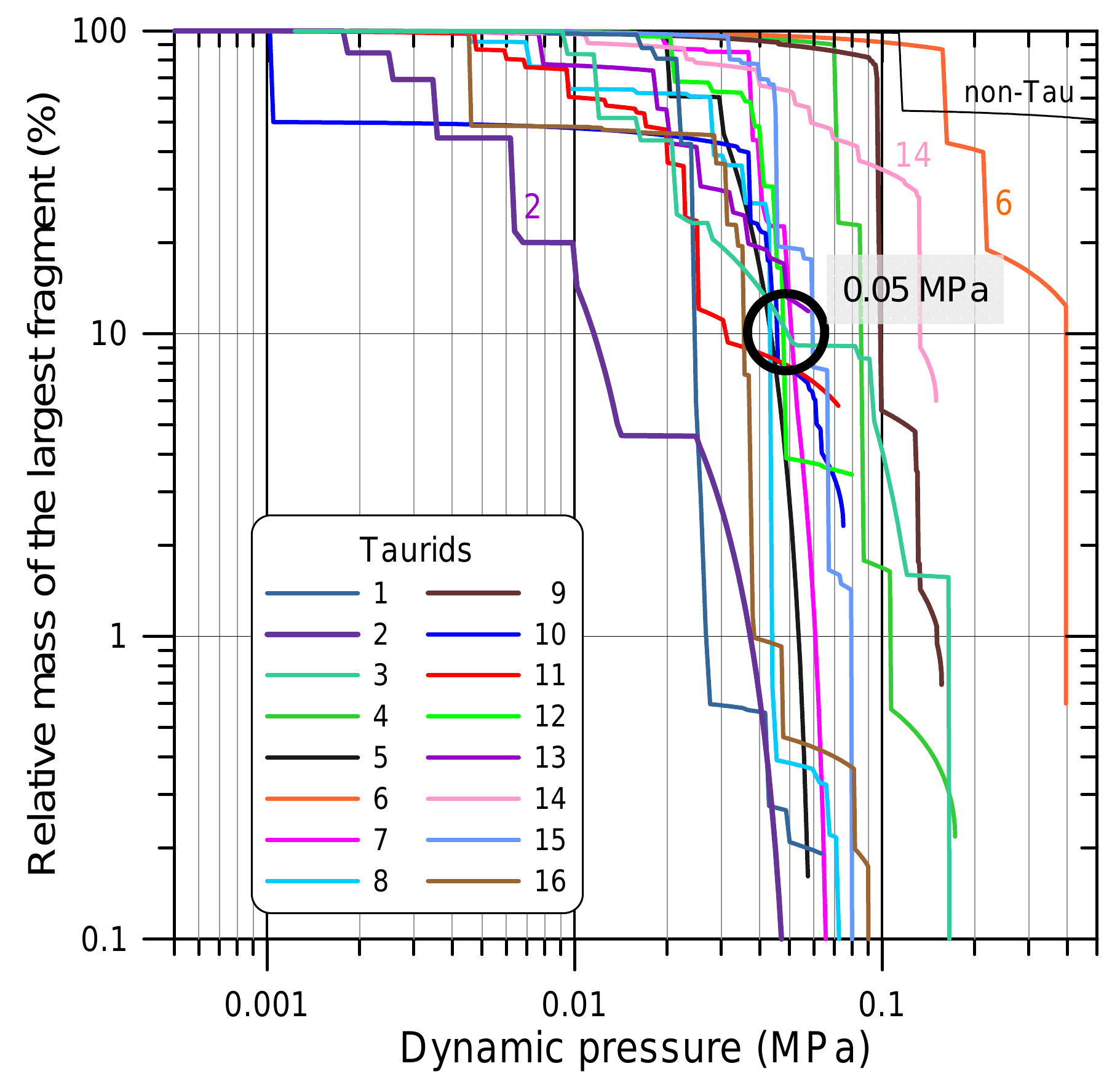}
\caption{The same as Fig.~\ref{press-mass} but with mass in relative scale.}
\label{press-mass-pct}
\end{figure}

Figure~\ref{press-mass-pct}  shows the same data as Fig.~\ref{press-mass}  but the fragment mass is plotted in relative scale as
a fraction of the initial mass. It is noticeable that many Taurid meteoroids reached the level of 10\% at nearly the same
pressure of about 0.05 MPa.  We therefore define the pressure $p_{90\%}$ as the dynamic pressure when the meteoroid lost
90\% of its initial mass.  It was in the range 0.04--0.06 MPa for 8 Taurids and within 0.025--0.1 MPa for 13 of the 16 Taurids. 
It therefore seems that a characteristic property of most Taurids is that 90\% of mass is lost, within a factor of two,
at a pressure of 0.05 MPa. The obviously deviating case was the huge meteoroid no.\ 2 with $p_{90\%}\sim 0.01$ MPa.
At 0.05 MPa more than 99.9\% of mass was lost. 
The opposite cases were small meteoroids no.\ 6 and 14 with $p_{90\%}$ 0.4 MPa and 0.13 MPa, 
respectively.

\begin{figure}
\centering
\includegraphics[width=9.1cm]{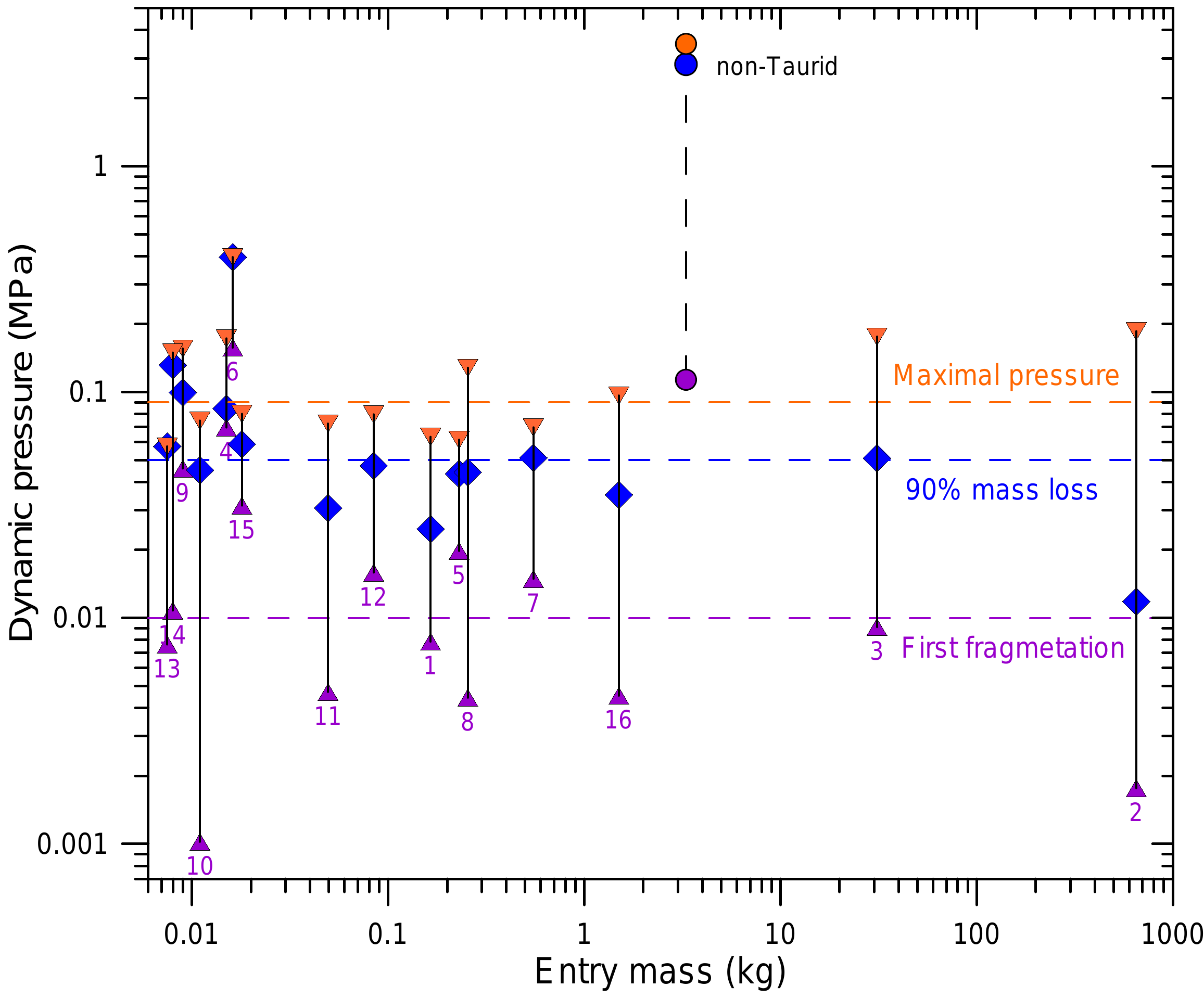}
\caption{The dynamic pressure at the first fragmentation, at the time when 90\% of mass was lost (i.e.\ the mass
of the largest fragment decreased to 10\% of the initial mass), and the maximal reached dynamic pressure as a function
of initial mass. 
For meteoroid numbers see Table~\ref{table}.
Median values are shown as horizontal dashed lines. The values for the non-Taurid are plotted for
comparison.}
\label{pressures}
\end{figure}

Figure~\ref{pressures} shows the characteristic pressures $p_1$, $p_{90\%}$, and $p_{\rm max}$ as a function of meteoroid mass.
Here $p_1$ is the pressure of first fragmentation and $p_{\rm max}$ is the maximal reached pressure. The values
are also listed in Table~\ref{table}. Figure~\ref{pressures} shows also the median values of the pressures. For the first fragmentation,
the median is 0.01 MPa, but there is very large scatter of the actual values. Meteoroid no.\ 10 broke up already at 0.001 MPa,
although it was a rather quiet splitting into two similarly sized fragments and the next fragmentation (probably of the smaller
of the two fragments) occurred at 0.01 MPa.  The huge meteoroid no.\ 2 suffered complete disintegration
between 0.0016 -- 0.01 MPa. On the other hand, some small meteoroids were much stronger. 
As mentioned above, meteoroid no.\ 6 did not break up
until 0.16 MPa.

The values of $p_{90\%}$, with median of 0.05 MPa, were already discussed. The median value of $p_{\rm max}$ is 0.09 MPa.
For almost all Taurids including the largest one, maximum pressure was within a factor of two of this median. The only exception 
was the strongest Taurid no.\ 6 with $p_{\rm max}= 0.4$ MPa. But the non-Taurid was even much stronger with $p_{\rm max}= 3.5$ MPa.

There is a weak trend of decrease of $p_1$ and $p_{90\%}$ with increasing meteoroid mass. Formally, it can be fitted by power index
of about $-0.16$. But there is huge scatter of individual values. Except for the largest meteoroid no.\ 2, the power law function does not provide
better prediction than using the median value. 

There is no trend for $p_{\rm max}$. Maximal pressure was usually reached at the last fragmentation point. It therefore expresses
the strength of the strongest macroscopic material contained in the meteoroid. Except for meteoroid no.\ 6, the strongest material 
seems to be nearly the same,
irrespective of the initial meteoroid size. In four cases (fireballs nos.\ 4, 8, 10, and 14), the maximal pressure was reached
shortly after the last fragmentation, when the largest surviving fragment was decelerated.  In these cases, the maximal pressure 
expresses just the lower limit of the strength of the last fragment. But all these fragments were small, with masses of 0.1 -- 0.7 g.

\subsection{Strength distribution}

Since the model provided the dynamic pressure and the amount of mass lost for each fragmentation, we can study the strength distribution inside the meteoroids. The strength of the lost mass is considered
to be equal to the dynamic pressure, at which it was lost. The statistics of strength was created for all 16 Taurids.
The mass lost by ablation was ignored in the statistics.

\begin{figure}
\centering
\includegraphics[width=8.5cm]{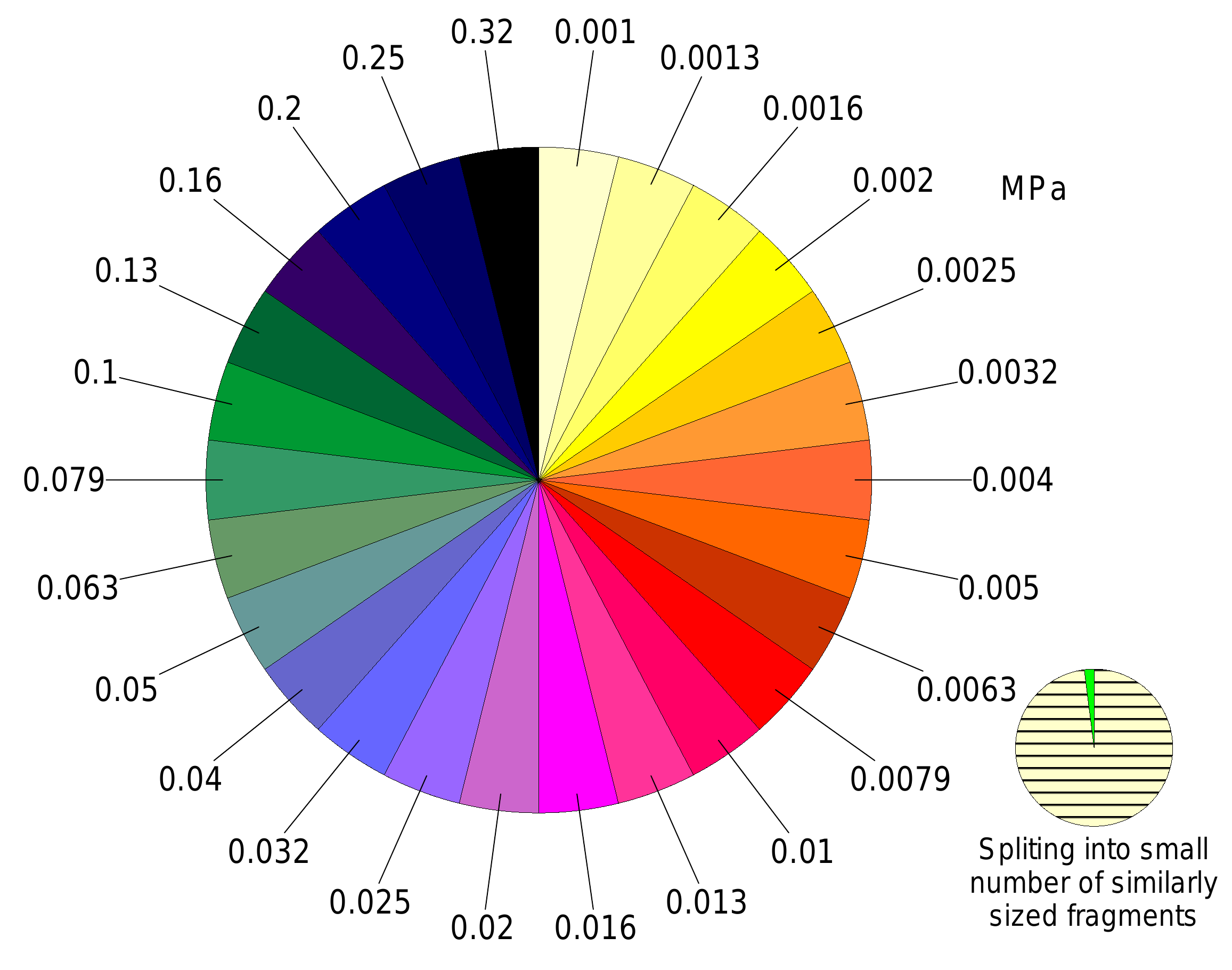}
\caption{Color coding of strengths in the pie charts in Fig.~\ref{pies}. The color scale is logarithmic. Three steps correspond
to the increase of strength two times.}
\label{piecolor}
\end{figure}

We used the color codes presented in Fig.~\ref{piecolor} to display the derived strengths graphically. The lowest strengths
are in yellow and orange, the highest in green and dark blue. If the mass was lost be splitting into small number of similarly
sized regular fragments rather than dust, eroding fragment or large number of small fragments, the color is hatched. 

\begin{figure*}
\centering
\includegraphics[width=14cm]{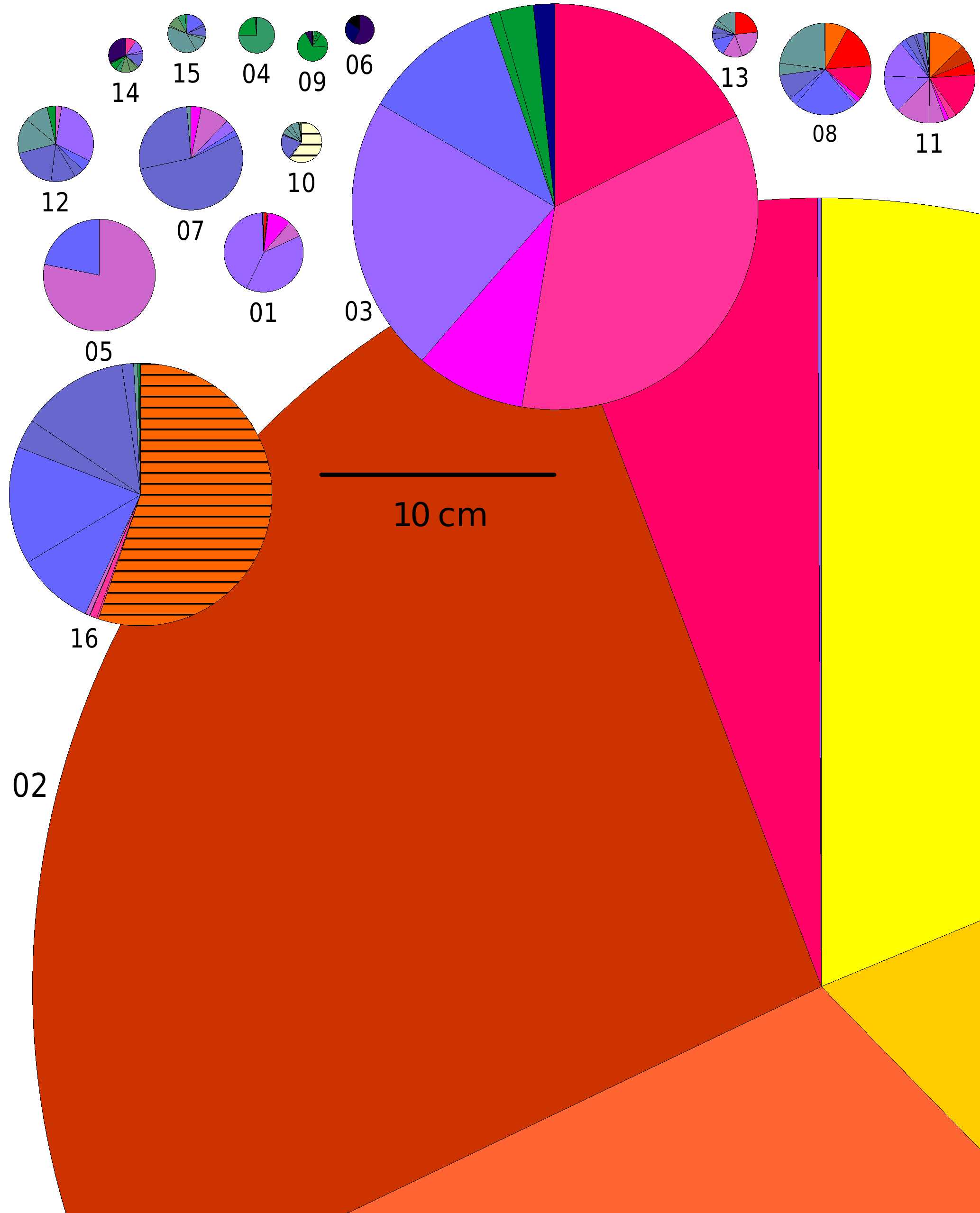}
\caption{Distribution of strength in 16 Taurid meteoroids. For color code see Fig.~\ref{piecolor}. Individual meteoroids
are represented by circles proportional to the meteoroid size (computed from the initial mass and density) and labeled by 
the numbers from Table~\ref{table}. The plotted fractions are fractions of released mass.}
\label{pies}
\end{figure*}

The results are presented in Fig.~\ref{pies}. It can be seen that the largest meteoroid no.\ 2 had very low strength. Only
tiny fraction, hardly visible in Fig.~\ref{pies}, had strength larger than 0.02 MPa. Material of strength higher than 0.1 MPa
represented $<0.01$\% of the mass and is invisible in the plot. The second largest meteoroid no.\ 3 was considerably
stronger. Its weakest part was as strong as the strongest 5\% of meteoroid no.\ 2. It also contained $\sim 5$\% of
material stronger than 0.08 MPa and 1.5\% stronger than 0.1 MPa. If we do not count splitting into similar pieces,
smaller meteoroids rarely contained the very weak material which formed 95\% of meteoroid no.\ 2.
Some medium-sized meteoroids like nos.\ 8 and 11 were very heterogeneous and contained the very weak material 
in small amounts.  Other medium-sized meteoroids like nos. 1, 5, and 7 were more homogeneous and
contained mostly material of medium strength of 0.02--0.03 MPa. The smallest meteoroids nos.\ 4, 9, 14, 15
were composed mostly from stronger material of strengths of 0.05--0.1 MPa; no.\ 6 even 0.15--0.4 MPa.

\begin{figure}
\centering
\includegraphics[width=8.5cm]{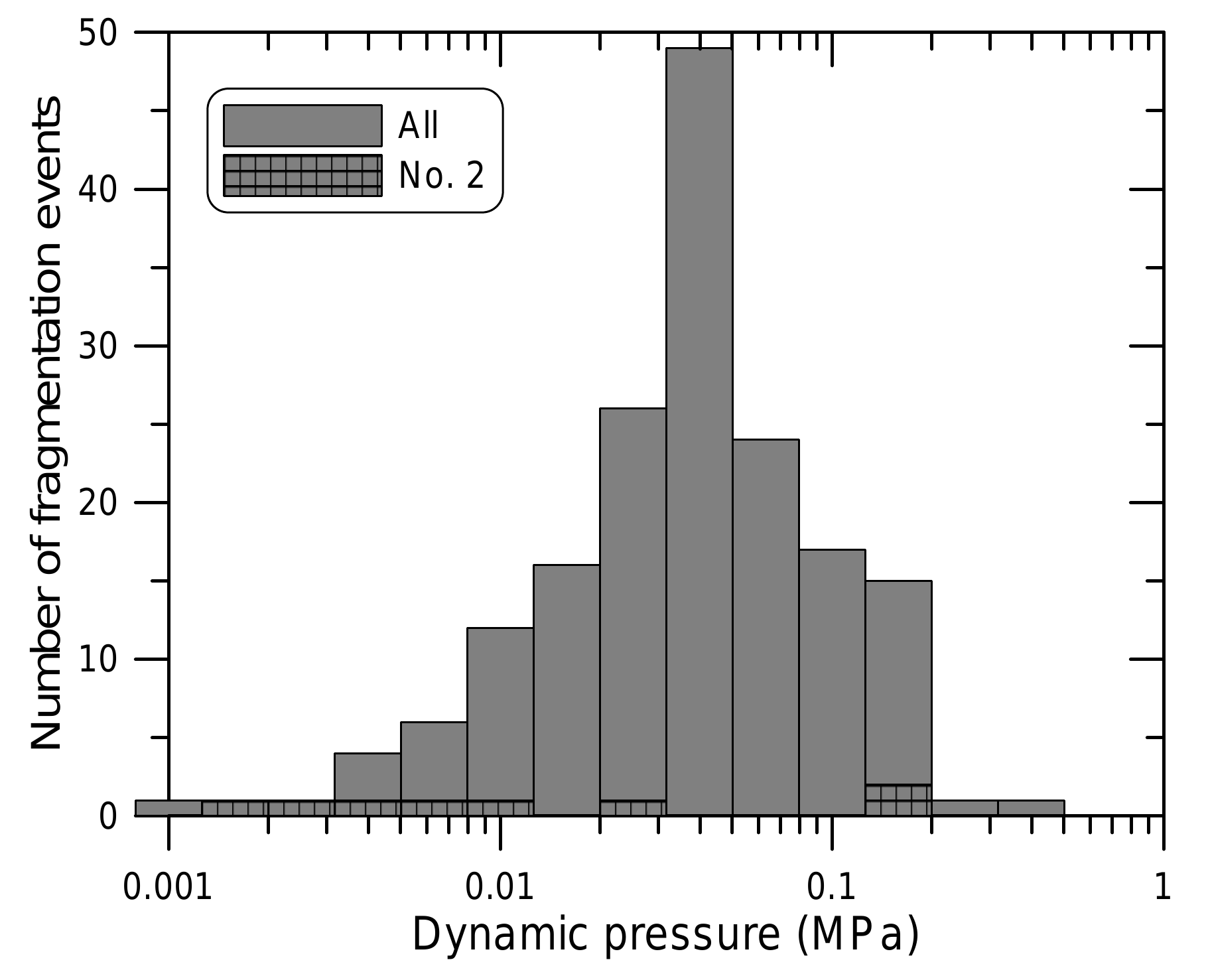}
\caption{Number of all fragmentation events for all 16 Taurids as a function of dynamic pressure. Fragmentations
of the largest meteoroid no.\ 2 are highlighted.}
\label{histog1}
\end{figure}

\begin{figure}
\centering
\includegraphics[width=8.5cm]{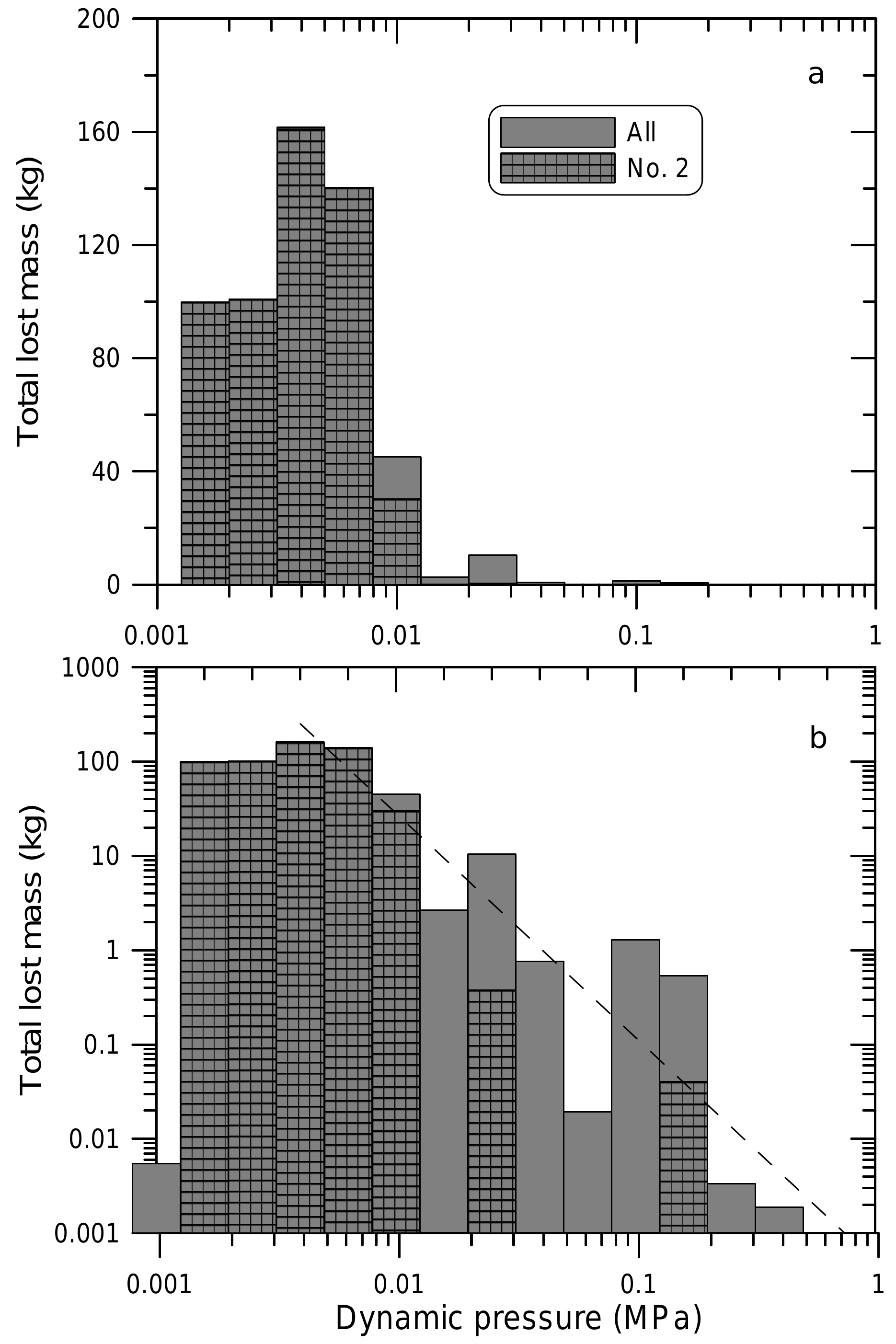}
\caption{Total mass lost in all fragmentation events for all 16 Taurids as a function of dynamic pressure. Fragmentations
of the largest meteoroid no.\ 2 are highlighted. The same data are shown in linear (a) and logarithmic (b) scale.}
\label{histog2}
\end{figure}

Figure~\ref{histog1} shows the histogram of fragmentation pressures. The largest number of fragmentation events
occurred at a pressure of about 0.04 MPa. Note that in contrast to Figs.~\ref{press-mass} and \ref{pies},
all modeled fragmentations, not only those of the largest surviving fragment, are considered in Fig.~\ref{histog1}.
From 174 total fragmentations, 159, i.e.\ more than 90\%, occurred at pressures 0.01--0.2 MPa. This statistics,
however, does not consider how representative individual fragmentations were, i.e.\ how large amount of mass was lost.
The distribution considering the lost mass is shown in Fig.~\ref{histog2}. Here the statistics is dominated by the four first
fragmentations of the largest meteoroid no.\ 2, which all occurred at pressures $<0.01$ MPa and  released huge amount of mass.
If we therefore  consider the Taurid material as a whole, the large majority had strengths 0.0015--0.008 MPa.
This strength is much lower than the tensile strength of snow, which is about 0.01 -- 0.04 MPa \citep{snow}.
The amount of stronger material decreased nearly exponentially with increasing strength (Fig.~\ref{histog2}b). 

\begin{figure}
\centering
\includegraphics[width=8cm]{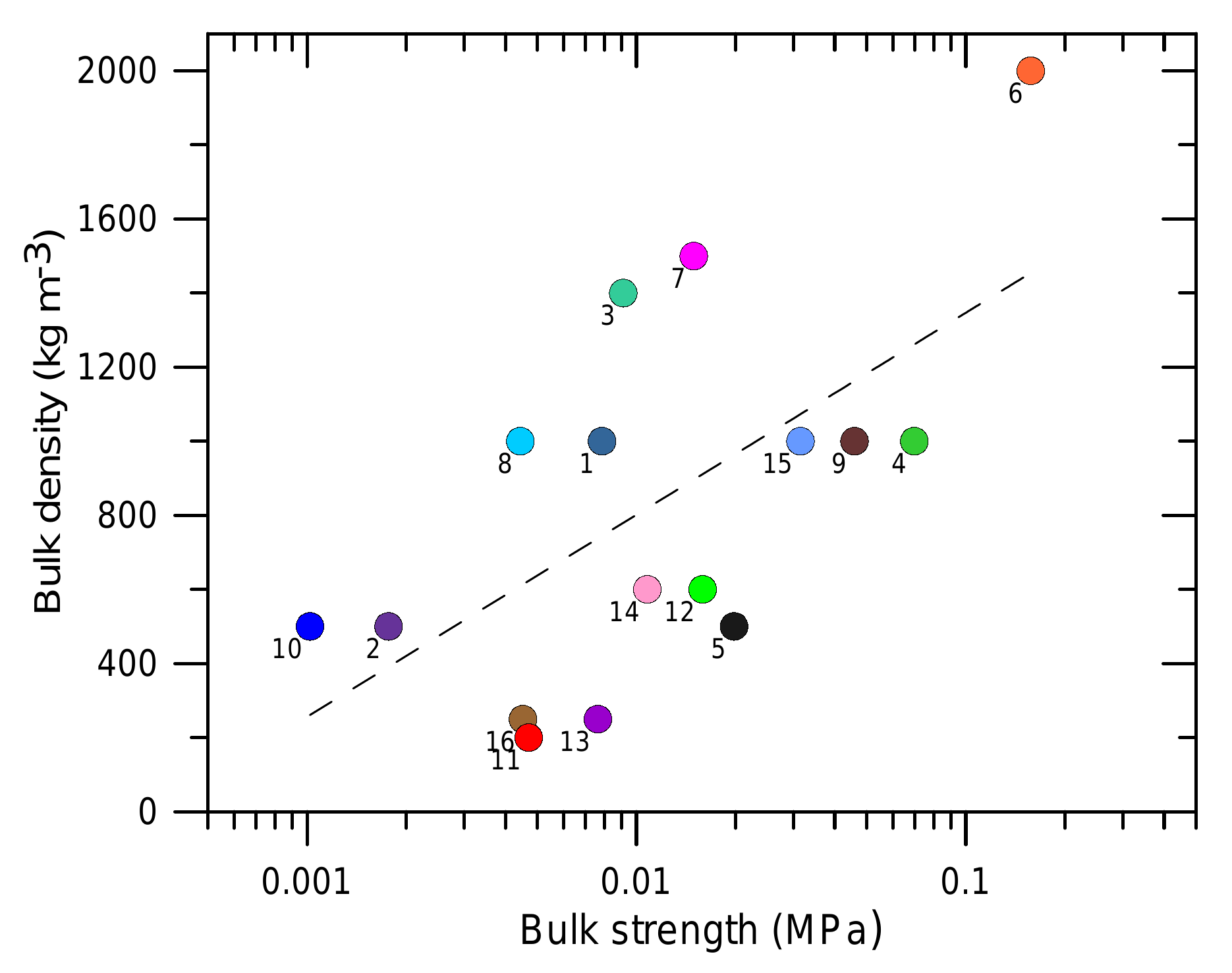}
\caption{Bulk densities of Taurid meteoroids as a function of dynamic pressure at the first fragmentation. 
For fireball numbers see Table~\ref{table}.}
\label{density}
\end{figure}

\subsection{Bulk densities}

Bulk densities of the incoming meteoroids are much more difficult to derive than fragmentation strengths.
As explained in Section~\ref{model}, the derived densities may be affected by unknown shapes of the
meteoroids.             
The computed densities are listed in Table~\ref{table} and cover wide range from 200 to 2000 kg m$^{-3}$.
The strongest meteoroid no.\ 6 had the highest density, but otherwise no clear correlation with strength was found.
There is only a weak correlation between the dynamic pressure at the first fragmentation, $p_1$ (which can be called
bulk strength), and bulk density (Fig.~\ref{density}). 
The density of the weakest and largest meteoroid no.\ 2 was estimated to 500 kg m$^{-3}$ but the uncertainty is
large in this case since steady-state ablation was not reached before the start of fragmentation. 
Note that the density of comet 2P/Encke was found to be rather uncertain $800 \pm 800$  kg m$^{-3}$ \citep{Sosa}.
The mean value for all comets  studied so far is $480 \pm 220$  kg m$^{-3}$ \citep{Groussin}.

\section{Discussion and conclusions}

We provided the most detailed study so far of the material properties of Taurid
meteoroids. It was important that we had a compact orbital sample and there
was no doubt that all studied meteoroids belonged to the Taurid stream.
Most of them belonged to the resonant swarm, which, in fact, could be described
in detail thanks to the quality of our orbital data \citep{Spurny}.

This study was based on the modeling of atmospheric fragmentation of meteoroids. 
The availability of precise radiometric light curves with high temporal resolution
was crucial for the modeling. It was also very important that our sample contained 
the extremely bright ($-18$ mag) Taurid fireball EN\,311015\_180520, which was
caused by a meteoroid approaching size of one meter. One of main results of our study
is that large Taurid meteoroids have different properties than smaller ones (cm-sized).
Note that Taurid stream is unique among all meteoroid streams in containing very
large meteoroids \citep{Brown}. We believe that EN\,311015\_180520 was not exceptional
and its properties are typical for Taurids of similar sizes. As mentioned in \citet{Spurny},
all probable Taurid superbolides detected by the US Government Sensors reached their maxima
at very high altitudes.

Our analysis showed that when taking the Taurid material as a whole, the majority has
very low strength of $<0.01$ MPa. 
Such low strength material exists mostly as large bodies 
($\gg10$ cm), can form a significant part of medium-sized bodies ($\sim 10$ cm),
and only rarely is encountered in smaller bodies, where it forms a minority.
Stronger materials exist in Taurids as well, but their occurrence decreases exponentially
with strength. Strong material forms small inclusions in large bodies 
or exists as small (cm-sized) separate bodies. A typical not too large Taurid lose 90\% of its mass when
the dynamic pressure reaches 0.05 MPa. All these properties suggest that it is unlikely that Taurids can produce
meteorites. The material, which could potentially be strong enough to
survive the atmospheric passage, is too small.

Taurid properties strongly suggest their cometary origin. Low strength and low density
(likely $<1000$ kg m$^{-3}$ for most material) is typical for comets. Even the asteroids
such as 2015 TX24 that are part of the Taurid new branch \citep{Spurny} are likely
of cometary origin. The existence of large fragile bodies within the stream suggest
a quiet process of comet disintegration. 
Some of these large bodies may have collided with the Earth in the past and some may do it in the future.
The grand comet hypothesis \citep{Clube} seems to be plausible.

\subsection*{Acknowledgments}
This work was supported by grant no.\ 19-26232X from the Czech Science Foundation.
The observations by DAFO were made possible thanks to the Praemium Academiae from the
Czech Academy of Sciences. The intitutional  project was  RVO:67985815.





\end{document}